\documentclass[twocolumn,showpacs]{revtex4}
\usepackage{eurosym}
\usepackage{graphicx}
\usepackage{dcolumn}
\usepackage{bm}
\usepackage{amsmath}
\usepackage{amsfonts}
\usepackage{amssymb}

\providecommand{\U}[1]{\protect\rule{.1in}{.1in}}

\begin{document}

\title{Sub-ballistic behavior in quantum systems with L\'evy noise}
\author{A. Romanelli}
\altaffiliation[Corresponding author.\\
]{\textit{E-mail address:} alejo@fing.edu.uy}
\author{R. Siri}
\author{V. Micenmacher}
\affiliation{Instituto de F\'{\i}sica, Facultad de Ingenier\'{\i}a\\
Universidad de la Rep\'ublica\\
C.C. 30, C.P. 11000, Montevideo, Uruguay}
\date{\today }

\begin{abstract}
We investigate the quantum walk and the quantum kicked rotor in
resonance subjected to noise with a L\'evy waiting time
distribution. We find that both systems have a sub-ballistic wave
function spreading as shown by a power-law tail of the standard
deviation.
\end{abstract}

\maketitle

%
In the last decades the study of simple quantum systems, such as the quantum
kicked rotor (QKR) \cite{Izrailev} and the quantum walk (QW) \cite{Kempe},
have exposed unexpected behaviors that suggest new challenges both
theoretical and experimental in the field of quantum information processing
\cite{Chuang}. The behavior of the QKR has two characteristic modalities:
dynamical localization (DL) and ballistic spreading of the variance in
resonance. These different behaviors depend on whether the period of the
kick is a rational or irrational multiple of $4\pi $. For rational multiples
the behavior of the system is resonant and the average energy grows
ballistically and for irrational multiples the average energy of the system
grows, for a short time, in a diffusive manner and afterwards DL appears.
Quantum resonance is a constructive interference phenomena and DL is a
destructive one. The DL\ and the ballistic behavior have already been
observed experimentally \cite{Moore,Kanem}. On the other hand the concept of
QW introduced in \cite{Aharonov,Godoy} is a counterpart of the classical
random walk. Its most striking property is its ability to spread over the
line linearly in time, this means that\ the standard deviation grows as $%
\sigma (t)\sim t$, while in the classical walk it grows as $\sigma (t)\sim
t^{1/2}$. We have developed \cite{alejo1,alejo2} a parallelism between the
behavior of the QKR and a generalized form of the QW showing that these
models have similar dynamics. In \cite{alejo3} we have investigated the
resonances of the QKR subjected to an excitation that follows an aperiodic
Fibonacci prescription; there we proved that the primary resonances retain
their ballistic behavior while the secondary resonances show a sub-ballistic
wave function spreading ($\sigma (t)\sim t^{c}$ with $0.5<c<1$) like the QW
with the same prescription for the coin \cite{Ribeiro}. Casati \textit{et}
\textit{al}. \cite{Casati} have studied the dynamics of the QKR kicked
according to a Fibonacci sequence outside the resonant regime, they found
sub-diffusive behavior for small kicking strengths and a threshold above
which the usual diffusion is recovered. More recently Schomerus and Lutz
\cite{Shomerus} investigated the QKR subjected to a L\'{e}vy noise \cite%
{Levy} and they show that this decoherence never fully destroys the DL of
the QKR but leads to a sub-diffusion regime for a short time before DL
appears.

In this article we investigate the QKR in resonant regime and the usual QW
when both are subjected to decoherence with a L\'{e}vy noise. In the case of
the QKR the model has two strength parameters whose action alternate in a
such way that the time interval between them follows a power law
distribution. In the case of QW the model uses two evolution operators whose
alternation follows the same power law distribution. We show that this noise
in the secondary resonances of the QKR and in the usual QW produces a change
from ballistic to sub-ballistic behavior. This change of behavior is similar
to that obtained for both systems when they are subjected to an aperiodic
Fibonacci excitation \cite{alejo3,Ribeiro}. %
%

\textit{L\'{e}vy distribution}.- We consider two time step unitary operators
$U_{0}$ and $U_{1}$ in a large sequence to generate the dynamics of the
quantum system. The time interval for the alternation of $U_{0}$ and $U_{1}$
is generated by a waiting time distribution $\rho (\Delta T)$, where $\Delta
T=iT$ \ with $T$ a time step and $i$ an integer. Then $i$ is the number of
times that $U_{0}$ is repeated before $U_{1}$ is applied once, \textit{e.g.}
the sequence of operators when the first interval is $\Delta T_{1}=4T$ and
the second is $\Delta T_{2}=2T$, is $%
U_{1}U_{0}U_{0}U_{1}U_{0}U_{0}U_{0}U_{0} $. In this paper we take $\rho
(\Delta T)$ in accordance with the L\'{e}vy distribution \ This distribution
appears frequently in nonlinear, fractal, chaotic and turbulent phenomena
\cite{Shlesinger,Klafter,Zaslavsky}, it includes a parameter $\alpha $, with
$0<\alpha \leq 2$, and it is identical to the Gaussian distribution when $%
\alpha =2$. When $\Delta T$ is large the asymptotic behavior of $\rho
(\Delta T)$ is $(1/\Delta T)^{1+\alpha }$, this implies that the second
moment of $\rho (\Delta T)$ is infinite when $\alpha <2$ and then there is
no characteristic size for the temporal jump, except in the Gaussian case.
It is just this absence of characteristic scale that makes L\'{e}vy random
walks scale-invariant fractals. As we are interested in the asymptotic
behavior of the QKR and QW, the most important characteristic of the L\'{e}%
vy noise is the power law shape of the tail. To capture the essence of the L%
\'{e}vy noise distribution, and simplify at the same time the numerical
calculation, we define the waiting time distribution as
\begin{equation}
\rho (t)=\frac{\alpha }{\left( 1+\alpha \right) T}\left\{
\begin{array}{cc}
1 & 0\leq t<T \\
\left( \frac{T}{t}\right) ^{\alpha +1} & t\geq T%
\end{array}%
\right. .  \label{Levy}
\end{equation}%
Then, the mechanism to obtain the time interval $\Delta T$ in agreement with
the previous discussion is the following: a) we sort a stochastic variable $%
\gamma $ with uniform distribution in $\left[ 0,1\right] $, b) we obtain $%
\xi $ as a solution of the equation $\gamma ={\int\limits_{0}^{\xi }{\rho (t)%
}dt}$ and finally c) $\Delta T=iT$ where $i$ is the integer part of $\xi $.
%
%

\textit{Quantum kicked rotor}.- The QKR Hamiltonian is
\begin{equation}
H=\frac{P^{2}}{2I}+K\cos \theta \sum_{n=1}^{\infty }\delta (t-nT),
\label{qkr_ham}
\end{equation}%
where the external kicks occur at times $t=nT$ with $n$ integer and $T$ the
kick period, $I$ is the moment of inertia of the rotor, $P$ the angular
momentum operator, $K$ the strength parameter and $\theta $ the angular
position. In the angular momentum representation, $P|\ell \rangle =\ell
\hbar |\ell \rangle $, the wave-vector is $|\Psi (t)\rangle =\sum_{\ell
=-\infty }^{\infty }a_{\ell }(t)|\ell \rangle $ and the average energy is $%
E(t)=\left\langle \Psi \right\vert H\left\vert \Psi \right\rangle
=\varepsilon \sum_{\ell =-\infty }^{\infty }\ell ^{2}\left\vert a_{\ell
}(t)\right\vert ^{2}$, where $\varepsilon =\hbar ^{2}/2I$. Using the Schr%
\"{o}dinger equation the quantum map is readily obtained from the
Hamiltonian (\ref{qkr_ham})
\begin{equation}
a_{\ell }(t_{n+1})=\sum_{j=-\infty }^{\infty }U_{\ell j}a_{j}(t_{n}),
\label{mapa}
\end{equation}%
where the matrix element of the time step evolution operator $U(\kappa )$ is
\begin{equation}
U_{\ell j}=i^{-(j-\ell )}e^{-ij^{2}\varepsilon T/\hbar }\,J_{j-\ell }(\kappa
),  \label{evolu}
\end{equation}%
$J_{m}$ is the $m$th order cylindrical Bessel function and its argument is
the dimensionless kick strength $\kappa \equiv K/\hbar $. The resonance
condition does not depend on $\kappa $ and takes place when the frequency of
the driving force is commensurable with the frequencies of the free rotor.
Inspection of eq.(\ref{evolu}) shows that the resonant values of the scale
parameter $\tau \equiv \varepsilon T/2\hbar $ are the set of the rational
multiples of $4\pi $, \textit{i.e.} $\tau =4\pi $ $p/q$. In what follows we
assume that the resonance condition is satisfied, therefore the evolution
operator depends on $\kappa $, $p$ and $q$. We call a resonance primary when
$p/q$ is an integer and secondary when it is not.

With the aim to generate the dynamics of the system we consider two values
of the strength parameter $\kappa _{1}$ and $\kappa _{2}$, and combine the
corresponding time step operators $U_{0}=U\left( \kappa _{1}\right) $ and $%
U_{1}=U\left( \kappa _{2}\right) $ in a large sequence. We have proved in
\cite{alejo3} that the ballistic behavior is maintained in the primary
resonances for any type of sequences of the operators $U_{0}$ and $U_{1}$
because the operators $U_{0}$ and $U_{1}$ commute. For the same reason the
antiresonance $p/q=1/2$ is not changed. Then we only need to study the
secondary resonances in this work.

\begin{figure}[th]
\begin{center}
\includegraphics[scale=0.3455]{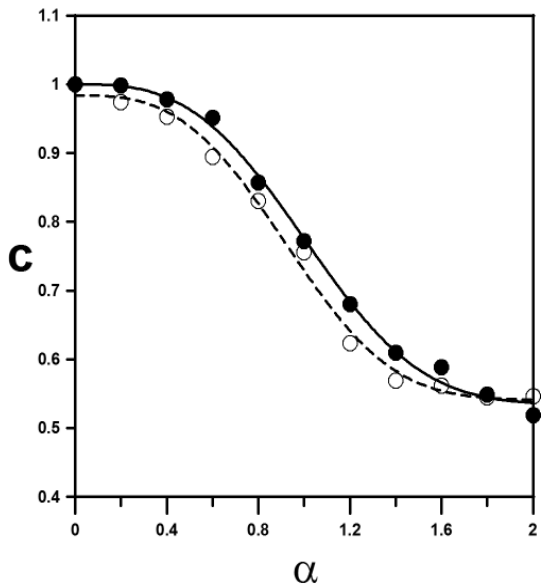}
\end{center}
\caption{The exponent $c$ of the power law of the standard deviation as a
function of the parameter $\protect\alpha $. (a) The black dots correspond
to the QKR with $\protect\kappa _{1}=1$, $\protect\kappa _{2}=-1$ and $%
p/q=1/3$, the full line is an adjustment. The parameters for the QKR\ are.;
(b) The white dots correspond to the QW with $\protect\theta _{1}=\protect%
\pi /3$, $\protect\theta _{2}=\protect\pi /6$, the dashed line is an
adjustment. }
\label{qkr3}
\end{figure}

Using the operators $U_{0}$, $U_{1}$ and the waiting time distribution (\ref%
{Levy}) we obtain the wave function spreading as given by the exponent $c$
of the asymptotic expression of the standard deviation $\sigma (t)=\sqrt{%
\sum_{\ell =-\infty }^{\infty }\ell ^{2}\left\vert a_{\ell }(t)\right\vert
^{2}}\sim t^{c}$. The initial condition for the wave-vector is the position
eigenstate $|0\rangle $, that is $a_{0}(0)=1$. The average standard
deviation $\sigma (t)$ is numerically obtained running a code with an
ensemble of $10000$ trajectories for several thousands of $T$. The table~(%
\ref{tabla})%
\begin{equation}
\begin{array}{ccccc}
p/q & 1/3 & 1/4 & 1/5 & 2/5 \\
c & 0.87 & 0.92 & 0.59 & 0.89%
\end{array}%
,  \label{tabla}
\end{equation}%
shows that $c$ depends on the ratio $p/q$. It is calculated with $\kappa
_{1}=0.5$, $\kappa _{2}=-0.5$ and $\alpha =1$. The value of $c$ remains
unchanged when $p/q$ is changed for $\left( q-p\right) /q$; this symmetry in
$c$ being a consequence of the trivial symmetry of the time step evolution
operator $U$ as was shown in \cite{alejo3}.
\begin{figure}[th]
\begin{center}
\includegraphics[scale=0.38]{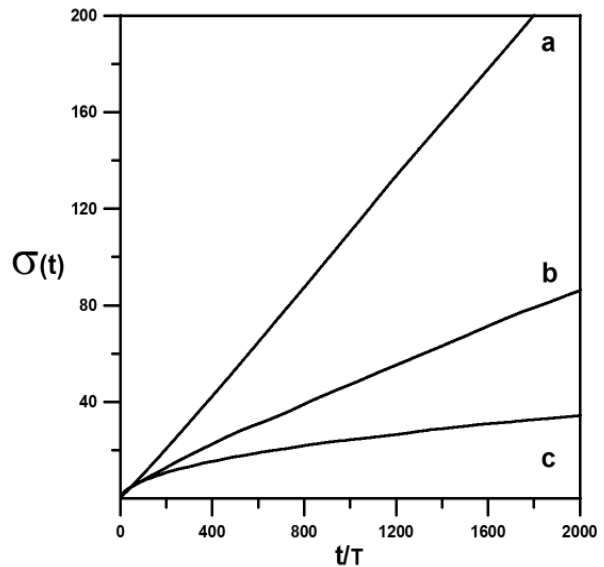}
\end{center}
\caption{The standard deviation for the QKR as a function of time in units
of $T$, with the parameters $\protect\kappa _{1}=1$, $\protect\kappa _{2}=-1$
and $p/q=1/3$. (a) $\protect\alpha =0.2$ and $c=0.998$; (b) $\protect\alpha %
=1$ and $c=0.772$; (c) $\protect\alpha =2$ and $c=0.518$. }
\label{qkr2}
\end{figure}
We have verified that the exponent $c$ has a dependence with the strength
parameters $\kappa _{1}$and $\kappa _{2}$ and its range is always between $%
0.5$ and $1$, thus the sub-ballistic behavior is maintained. In Fig.~\ref%
{qkr2} $\sigma (t)$ is plotted for several values of the parameter $\alpha $%
, displaying the qualitative differences between the periodic case $(\alpha
\sim 0,$ $c\sim 1)$, the L\'{e}vy noise case $(0<\alpha <2$ and $0.5<c<1)$
and the Gaussian case $(\alpha \sim 2,$ $c\sim 0.5)$. The exponent $c$ is
plotted in Fig.~\ref{qkr3} as a function of $\alpha $ showing a clear
dependence of $c$ with $\alpha $. We have also studied higher moments of
order four and six, they all have smaller exponents than those obtained with
a periodical sequence, thus the asymptotic behavior of these moments is
consistent with the power-law behavior of the second moment.

\textit{Quantum walk}.- 
\begin{figure}[th]
\begin{center}
\includegraphics[scale=0.38]{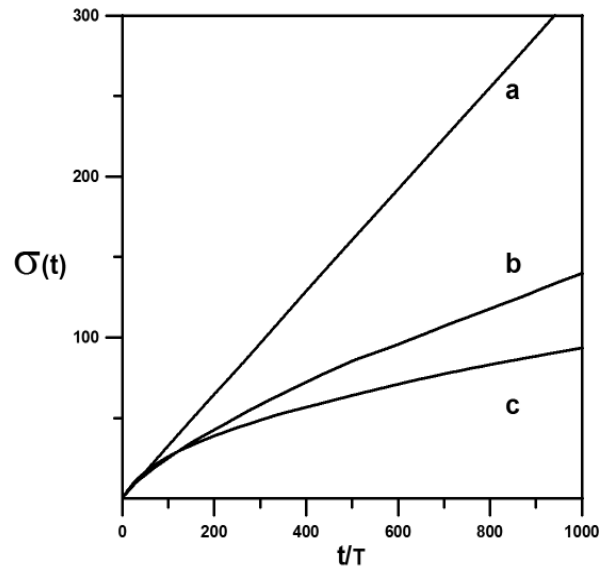}
\end{center}
\caption{The standard deviation for the QW as a function of time in units of
$T$ with the parameters $\protect\theta _{1}=\protect\pi /3$ and $\protect%
\theta _{2}=\protect\pi /6$. (a) $\protect\alpha =0.2$ and $c=988$; (b) $%
\protect\alpha =1$ and $c=0.71$; (c) $\protect\alpha =2$ and $c=0.546$. }
\label{qw1}
\end{figure}
\label{walked}The standard QW corresponds to a one-dimensional evolution of
a quantum system (the walker), in a direction which depends on an additional
degree of freedom, the chirality, with two possible states:
\textquotedblleft left\textquotedblright\ $|L\rangle $\ or \textquotedblleft
right\textquotedblright\ $|R\rangle $. The global Hilbert space of the
system is the tensor product $H_{s}\otimes H_{c}$ where $H_{s}$ is the
Hilbert space associated to the motion on the line and $H_{c}$ is the
chirality Hilbert space. Let us call $T_{-}$ ($T_{+}$) the operators in $%
H_{s}$ that move the walker one site to the left (right), and $|L\rangle
\langle L|$ and $|R\rangle \langle R|$ the chirality projector operators in $%
H_{c}$. We consider the unitary transformations
\begin{equation}
U(\theta )=\left\{ T_{-}\otimes |L\rangle \langle L|+T_{+}\otimes |R\rangle
\langle R|\right\} \circ \left\{ I\otimes K(\theta )\right\} ,  \label{Ugen}
\end{equation}%
where $K(\theta )=\sigma _{z}e^{-i\theta \sigma _{y}}$, $I$ is the identity
operator in $H_{s}$, and $\sigma _{y}$ and $\sigma _{z}$ are Pauli matrices
acting in $H_{c}$. The unitary operator $U(\theta )$ evolves the state in
one time step, $|\Psi (t+1)\rangle =U(\theta )|\Psi (t)\rangle $. Here we
are generalizing the QW to the case where different quantum coins are
applied as in \cite{Ribeiro}. As for the QKR, we combine two different step
operators $U_{0}=U(\theta _{1})$ and $U_{1}=U(\theta _{2})$, with $\theta
_{1}\neq \theta _{2}$, into a large sequence where we apply the same L\'{e}%
vy waiting time distribution. The wave function spreading is given by the
exponent $c$ in $\sigma (t)=\sqrt{\sum_{i=0}^{\infty }i^{2}\left( \left\vert
a_{i}(t)\right\vert ^{2}+\left\vert b_{i}(t)\right\vert ^{2}\right) }\sim
t^{c}$. We take as the initial condition for the QW the position eigenstate $%
|0\rangle $, with chirality $(1,0)$ for the calculations of Fig.~\ref{qw1}
and Fig.~\ref{qkr3}, and $(1,i)/\sqrt{2}$ for the calculations of Fig.~\ref%
{qw3}. The standard deviation $\sigma (t)$ is numerically obtained for an
ensemble of $1000$ trajectories for each value of the parameter $\alpha $ of
the L\'{e}vy distribution. In Fig.~\ref{qw1} $\sigma (t)$ is plotted for
different values of $\alpha $ and the sub-ballistic behavior is clearly
shown. In Fig.~\ref{qkr3} $c$ is plotted as a function of the parameter $%
\alpha $, here a functional dependence between them is evident. In Fig.~\ref%
{qw3} $c$ is plotted as a function of $\theta $, where $\theta =\theta
_{1}=-\theta _{2}$This figure shows that the range of $c$ is always $\left[
0.5,1\right] $, thus the sub-ballistic behavior is independent of the value
of $\theta $ (except for the trivial cases $\theta =0$, $\theta =\pi /2$);
additionally this figure is in concordance with Fig. $3$ in ref. \cite%
{Ribeiro} These figures show qualitative similarities with the corresponding
figures for the QKR, and point to the parallelism between the QW and the QKR
in the secondary resonance regime. Again in this system, the moments of
order four and six are consistent with the power-law behavior of the second
moment.
\begin{figure}[th]
\begin{center}
\includegraphics[scale=0.3455]{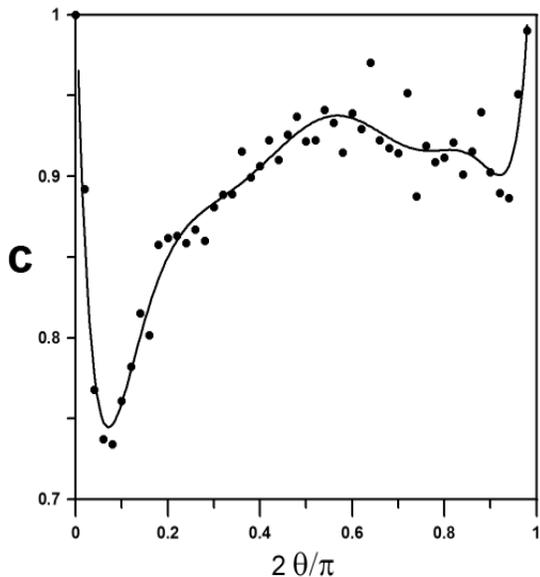}
\end{center}
\caption{The exponent $c$ that characterizes de power law of the standard
deviation for the QW as a function of the parameter $\protect\theta $. The
evolution is obtained with $\protect\theta =\protect\theta _{1}=-\protect%
\theta _{2}$ in the interval $\left[ 0,\protect\pi /2\right) $. The
dots correspond to the calculation and the full line is a polynomial
adjustment} \label{qw3}
\end{figure}
%
%
\textit{Conclusion}.- The quantum resonances and the DL of the QKR\ have
been experimentally observed in samples of cold atoms interacting with a
far-detuned standing wave of laser light \cite{expwave,expqkr,Ammann} and in
particular the secondary resonances have been recently observed by Kanem et
al. \cite{Kanem}. On the other hand several systems have been proposed as
candidates to implement the QW model. These proposals include atoms trapped
in optical lattices \cite{Dur}, cavity quantum electrodynamics \cite{Sanders}
and nuclear magnetic resonance in solid substrates \cite{Du,Berman}. All
these proposed implementations face the obstacle of decoherence due to
environmental noise and imperfections. Thus the study of the behavior of
these systems, subjected to different types of noise, is very important for
the design and construction of future technologies. Here we proposed the
study of the QKR and QW subjected to noisy pulses with a L\'{e}vy waiting
time distribution. As Gaussian noise is a particular case of the L\'{e}vy
noise, then our study is open to wider experimental situations. We prove
that for QKR\ and QW the L\'{e}vy noise does not break completely the
coherence in the dynamics but produces a sub-ballistic behavior in both
systems, as an intermediate situation between the ballistic and the
diffusive behavior. Then QKR and QW have essentially the same dynamical
evolution and our results fortify the previously established parallelism
between them \cite{alejo1,alejo2,alejo3}. More generally we can say that the
dynamical evolution of the QKR and QW show certain patterns that seem to be
common to a greater class of systems that are defined mainly by their
symmetries and not by their microscopic details. The existence of an
universality in the behavior of these systems suggests that one is allowed a
larger flexibility in the choice of the physical systems to build quantum
computers. It is important to highlight that the sub-ballistic behavior
obtained in this paper for the QW and the QKR is essentially the same as
that obtained for these systems when the perturbation follows a Fibonacci
prescription \cite{Ribeiro,alejo3}. The reason may lie in the fact that
behind the Fibonacci prescription hides the lack of a typical scale in the
sequence of the operators $U_{0}$ and $U_{1}$ which leads to a power law
\cite{Henrik} in the standard deviation, in the same way as for the L\'{e}vy
noise. \bigskip

We acknowledge the support from PEDECIBA and PDT S/C/OP/28/84.

\end{document}